\documentclass[a4paper,twoside,11pt,english,fleqn]{article}

\usepackage[intlimits]{amsmath}
\usepackage{amssymb,amsthm}
\usepackage{graphicx}
\usepackage[dvips]{epsfig}
\usepackage{babel}
\usepackage[T1]{fontenc}
\usepackage{fancyhdr,stmaryrd}
\usepackage{color}
\usepackage{times,euler,euscript}
\usepackage[modulo]{lineno}  

\DeclareMathAlphabet{\mathscr}{T1}{pzc}{m}{it}

  	\newtheorem{theorem}{Theorem}[subsection]
  	
  	\newtheorem{lemma}[theorem]{Lemma}
  	\newtheorem{proposition}[theorem]{Proposition}
\theoremstyle{definition}
	\newtheorem{definition}[theorem]{Definition}
	\newtheorem{notation}[theorem]{Notation}
	\newtheorem{remark}[theorem]{Remark}
	\newtheorem{example}[theorem]{Example}

\fancypagestyle{plain}{
  \fancyhf{}\fancyfoot[LC,RC]{}
  \fancyfoot[LE,RO]{$\thepage$}
  
  }

\linenumbers

\DeclareMathOperator{\id}{Id}
\renewcommand{\phi}{\varphi}
\renewcommand{\epsilon}{\varepsilon}
\newcommand{\fl}{\to}
\newcommand{\dfl}{\Rightarrow}
\newcommand{\tfl}{\Rrightarrow}
\newcommand{\mon}[1]{\langle#1\rangle}

\newcommand{\roundup}[1]{\left\lceil#1\right\rceil}
\newcommand{\rounddown}[1]{\left\lfloor#1\right\rfloor}
\newcommand{\ens}[1]{\{#1\}}
\newcommand{\Nb}{\mathbb{N}}

\renewcommand{\Pr}{\EuScript{P}}

\newcommand{\figeps}[1]{\raisebox{-1.25mm}{\includegraphics{#1.eps}}}

\newcommand{\titre}{Polygraphs for termination of \\ left-linear term rewriting systems}
\newcommand{\jour}{\today}
\newcommand{\auteur}{Yves GUIRAUD}
\newcommand{\adresse}{INRIA Lorraine -- LORIA -- yves.guiraud@loria.fr}

\begin{document}

\thispagestyle{empty}

\noindent\MakeUppercase{\textbf{\Large \titre}} 

\strut\hfill\textbf{\jour}

\noindent{\textbf{\Large \auteur}}

\noindent{\adresse}

\vspace{5mm}
\begin{em}
\noindent\textbf{Abstract --} We present a methodology for proving termination of left-linear term rewriting systems (TRSs) by using Albert Burroni's polygraphs, a kind of rewriting systems on algebraic circuits. We translate the considered TRS into a polygraph of minimal size whose termination is proven with a polygraphic interpretation, then we get back the property on the TRS. We recall Yves Lafont's general translation of TRSs into polygraphs and known links between their termination properties. We give several conditions on the original TRS, including being a first-order functional program, that ensure that we can reduce the size of the polygraphic translation. We also prove sufficient conditions on the polygraphic interpretations of a minimal translation to imply termination of the original TRS. Examples are given to compare this method with usual polynomial interpretations.
\end{em}

\section{Introduction}
\label{Section:Introduction}

Termination is a fundamental property of rewriting systems, since it ensures that the rule-based computations they define end with a result~\cite{BaaderNipkow98}. Even if this is an undecidable property for a general term rewriting system (TRS), many different techniques have been developped for this purpose. Among them, we are particularly interested into polynomial interpretations~\cite{Lankford79}: indeed, when one can prove termination of a (first-order) functional program with a polynomial interpretation, there are many cases where one can deduce an implicit complexity bound for the function that the program computes~\cite{CichonLescanne92,BonfanteCichonMarionTouzet01}. However, polynomial interpretations on TRSs have limits and here we address two of them.

First, as explained in~\cite{BonfanteGuiraud06}, the interpretation of a term conveys a mixed up information, containing a common bound on the size of the values to be computed and the size of the computation itself. Thus, the idea is that the coefficients one computes are higher than necessary: this increases the time to find a correct polynomial interpretation and decreases the precision of the computational complexity bound. For example, let us consider the functional program $\ens{D(0) \fl 0, \: D(s(x))\fl s(s(D(x)))}$ computing the "double" function on natural numbers. Then one can prove that the lower polynomial interpretation yielding its termination takes $D$ to $P(D)(X)=3X$: this is the sum of the size $2X$ of the computed value and the number $X$ of rewriting steps required to reach it on an input of size $X$. With the help of dependency pairs~\cite{ArtsGiesl00}, one can lower the interpretation of $D$ to $2X$. It is possible that, by application of several other methods, one could prove that $D$ can be interpreted to the polynomial $X$. But, atop of the complication of the process, we are not sure that theoretical results exist to state that this is an implicit complexity bound for the double function.

The second limit we consider comes with TRSs that do not admit simplification orders. The functional program $\ens{M(0,x) \fl 0, \: M(x,0) \fl x, \: M(s(x),s(y)) \fl M (x,y), \: Q(0,x) \fl 0, \: Q(s(x),y) \fl s(Q(M(x,y),y))}$, computing division on natural numbers, is an example of this class. Indeed, in the last rule, if one replaces $y$ by $s(x)$, the left-hand side $l(x,s(x))$ can be embedded into the right-hand side $r(x,s(x))$: hence, for any simplification order $>$, we have $r(x,s(x))>l(x,s(x))$ while proving termination with this strict order would require the reverse strict inequality. Nonetheless, this TRS is terminating and this can be proved, for example, by using dependency pairs, then semantic labelling and, finally, some simplification order. As above, this means that it is more complicated to prove termination of these systems and that we do not know if we can deduce an implicit complexity bound from this proof.

In order to solve these problems, we propose to use another formalism for expressing rewriting-based computations: \emph{higher-dimensional rewriting}~\cite{Lafont03}, a Turing-complete model~\cite{BonfanteGuiraud06}, based on Albert Burroni's polygraphs~\cite{Burroni93}, which could be described, at first glance, as an algebraic description of term graph rewriting systems~\cite{Plump99} and interaction nets~\cite{Lafont90}. Let us give the polygraphic versions of the two programs we have seen. For the double function, terms are replaced by $2$-dimensional algebraic circuits built upon the elementary gates~$\figeps{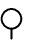}$ for $0$, $\figeps{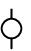}$ for $s$ and $\figeps{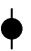}$ for $D$. The rewriting rules are replaced by the following ones:
\begin{center}\input{3-cellules-double.pstex_t}\end{center}

\noindent Concerning the division on natural numbers, we still use the gates $\figeps{cons-0}$ and $\figeps{cons-1}$, plus $\figeps{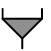}$ for $M$ and $\figeps{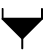}$ for~$Q$. We also need two extra gates $\figeps{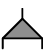}$ and $\figeps{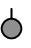}$ which are central in the present study and will be discussed later. The five rewriting rules are translated as follows:
\begin{center}\input{3-cellules-quotient.pstex_t}\end{center}

Polygraphs are higher-dimensional categories which are free in every dimensions. They have been introduced by Albert Burroni in order to provide a unified algebraic structure to many objects from theoretical computer science with an emphasis on rewriting systems. Yves Lafont has started the study of the computational properties of polygraphs~\cite{Lafont95,Lafont03}. Until now, polygraphs have been proved to unify several objects such as abstract, word and term rewriting systems~\cite{Guiraud06jpaa}, Petri nets~\cite{Guiraud06tcs} or formal proofs of propositional logics~\cite{Guiraud06apal}. 

We think that polygraphs are particularly suited for proving termination of TRSs using adaptated polynomial interpretations and, particularly, for functional programs. Indeed, we can see on the examples that, given a TRS, its associated polygraph is a quite direct translation, so that programs are written as polygraphs in a natural way. Furthermore, we have proved that the termination of the polygraph implies the termination of the rewriting system, provided it is left-linear~\cite{Guiraud06jpaa}. Finally, we have developped a tool called polygraphic interpretations~\cite{Guiraud06jpaa}, giving, on examples, some implicit complexity information which is much finer than the one we get on terms~\cite{BonfanteGuiraud06}: in the case of the double function, we get the $X$ bound we have discussed. The reason comes from the ability of polygraphic interpretations to differentiate functions from constructors in functional programs, as does the dependency pairs method.

However, the standard translation of a TRS into a polygraph generates a huge object, with many more rewriting rules: indeed, in the polygraphic framework, one has to explicitely handle duplications~$\figeps{delta}$, erasures~$\figeps{epsilon}$ and even permutations~$\figeps{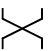}$, which means that one needs to add all the rules to compute these operations. This may have some advantages: for example, in the polygraphic setting, commutativity equations can be directed in a terminating way~\cite{Lafont03,Guiraud06jpaa}. But, for the moment, this expliciteness also has practical drawbacks. Indeed, there is only one result linking the termination of a TRS to the one of its standard polygraphic translation and this requires to consider all the rules of the polygraph, including the extra ones. And there can be many of them: for a term rewriting system with $m$ sorts, $n$ operations and $p$ rules, the standard polygraphic translation has $p+2n(m+1)+m(m^2+6m+5)$ rules. Even if we know that these extra rules have nice computational properties, including termination and confluence~\cite{Guiraud06jpaa}, we could not, until now, set aside some or all of these rules, thus making any practical use of the polygraphic method really hard, at best.

In order to correct this problem, we propose in this study several results that allow us to discard some or all of these extra rules in order to alleviate the computational burden they otherwise generate. We think that the results we prove here make it possible to automatically prove termination of TRSs by polygraphic interpretations and, for some functional programs, to give an implicit complexity bound at the same time. We plan to test such a prover-bounder on the Termination Problems DataBase (\emph{http://www.lri.fr/$\sim$marche/tpdb}): this will give essential information on the possible efficiency of our method compared to other ones, together with a general view on which systems it is most suited at and for which systems it can be improved. 

This paper is organized into two main sections, apart from this introduction and the conclusion. In section~\ref{Section:Polygraphs}, we recall the special case of polygraphs we consider (\ref{Subsection:Polygraphs}), then we explain the method of polygraphic interpretation (\ref{Subsection:Interpretations}) and how to translate a TRS into a polygraph (\ref{Subsection:Translations}). Section~\ref{Section:Results} contains the conditions for reducing the size of the polygraph one has to consider to prove termination of a left-linear TRS: theorem~\ref{Theorem:GeneralCase} can always be applied to discard a family of extra rules, theorem~\ref{Theorem:PlanarLinearCase} is a special case that allows one to consider no extra rules, theorem~\ref{Theorem:FunctionalProgramCase} is dedicated to the case of first-order functional programs and, finally, theorems~\ref{Theorem:NonDuplicating} and~\ref{Theorem:SpecialInterpretations} give sufficient conditions on a proof by polygraphic interpretation to discard several or all of the extra rules. 

For some basic notions of rewriting we do not recall, the reader can consult~\cite{BaaderNipkow98}. The author wishes to thank Frédéric Blanqui, Guillaume Bonfante, Yves Lafont and Philippe Malbos for many valuable discussions about polygraphic interpretations.

\section{Polygraphs, interpretations and term rewriting systems translations}
\label{Section:Polygraphs}

\subsection{Polygraphs}
\label{Subsection:Polygraphs}

The general definition of polygraph can be found in documents by Albert Burroni, Yves Lafont and François Métayer~\cite{Burroni93,Lafont03,Metayer03,Lafont06,LafontMetayer06}. Here we give a rewriting-minded presentation of a special case of polygraphs, seeing them as rewriting systems on algebraic circuits.

\begin{definition} \label{Definition:Polygraph}
A \emph{monoidal $\mathit{3}$-polygraph} is a composite object consisting of \emph{cells}, \emph{paths} and \emph{compositions} organized into \emph{dimensions}.

\emph{Dimension $\mathit{1}$} contains elementary sorts called \emph{$\mathit{1}$-cells} and represented by wires. Their concatenation~$\star_0$ yields product types called \emph{$\mathit{1}$-paths} and pictured as juxtaposed vertical wires. The empty product~$\ast$ is also a $1$-path, represented by the empty diagram. 

\emph{Dimension $\mathit{2}$} is made of operations called \emph{$\mathit{2}$-cells}, with a finite number of typed inputs and outputs. They are pictured as circuit gates, with inputs at the top and outputs at the bottom. Using all the $1$-cells and $2$-cells as generators, one builds circuits called \emph{$\mathit{2}$-paths}, using the following two compositions:
\begin{center}\input{2-compositions.pstex_t}\end{center}

\noindent The constructions are considered \emph{modulo} some relations, including topological deformation: one can stretch or contract wires freely, move $2$-cells, provided one does not create crossings or break wires. Each $2$-cell and each $2$-path $f$ has a $1$-path $s_1(f)$ as input, its \emph{$\mathit{1}$-source}, and a $1$-path $t_1(f)$ as output, its \emph{$\mathit{1}$-target}. The compact notation $f:s_1(f)\dfl t_1(f)$ summarizes these facts. 

\emph{Dimension $\mathit{3}$} contains rewriting rules called \emph{$\mathit{3}$-cells}. They always transform a $2$-path into another one with the same $1$-source and the same $1$-target. Using all the $1$-cells, $2$-cells and $3$-cells as generators, one can build reductions paths called \emph{$\mathit{3}$-paths}, by application of the following three compositions, defined for~$F$ going from $f$ to $f'$ and~$G$ going from $g$ to $g'$: $F\star_0 G$ goes from $f\star_0 g$ to $f'\star_0 g'$; when $t_1(f)=s_1(g)$, then $F\star_1 G$ goes from $f\star_1g$ to $f'\star_1 g'$; when $f'=g$, then $F\star_2 G$ goes from $f$ to $g'$. These constructions are identified \emph{modulo} some relations, given in~\cite{Guiraud06apal}, where their $3$-dimensional nature was explained. The relations allow one to freely deform the constructions in a reasonable way: in particular, they identify paths that only differ by the order of application of the same $3$-cells on non-overlapping parts of a $2$-path. A $3$-path is \emph{elementary} when it contains exactly one $3$-cell. Each $3$-cell and each $3$-path $F$ has a $2$-path $s_2(F)$ as left-hand side, its \emph{$\mathit{2}$-source}, and a $2$-path $t_2(F)$ as right-hand side, its \emph{$\mathit{2}$-target}. The notation $F:s_2(F)\tfl t_2(F)$ stands for these facts. 

In this study, \emph{polygraph} always means \emph{monoidal $\mathit{3}$-polygraph}. For polygraphs, rewriting notions are defined in a similar way as for TRSs, with terms replaced by $2$-paths, reduction steps by elementary $3$-paths and reduction paths by $3$-paths~\cite{Guiraud06jpaa}. Hence, a \emph{normal form} in a polygraph $\Pr$ is a $2$-path~$f$ which is the $2$-source of no elementary $3$-path. The polygraph $\Pr$ \emph{terminates} when it does not contain infinite families $(F_n)_{n\in\Nb}$ of elementary $3$-paths such that $t_2(F_n)=s_2(F_{n+1})$ for all $n$. Other rewriting properties, such as \emph{confluence} or \emph{convergence} are also defined in an intuitive way. If $X$ is a family of $i$-cells in $\Pr$, we denote by $\mon{X}$ the $i$-paths of $\Pr$ whose generating $i$-cells are all in $X$. If $i=3$ and if there exists a $3$-path in $\mon{X}$ from $f$ to $g$, we use the notation $f\tfl_X g$.When this path is elementary, we write $f\tfl_X^1 g$. If $X=\ens{\alpha}$, we write $f\tfl_{\alpha}g$. 
\end{definition}

\begin{example} \label{Example:Polygraph}
We have already seen two examples of polygraphs in the introduction. The following one computes the addition $\figeps{fonction-2-b}$ and the multiplication $\figeps{fonction-2}$ on natural numbers $\mon{\figeps{cons-0},\figeps{cons-1}}$, provided one adds the rules for the computation of~$\figeps{delta}$ and~$\figeps{epsilon}$, as we explain in~\ref{Subsection:Translations}:
\begin{center}\input{3-cellules-polynomes.pstex_t}\end{center}

\noindent Note that we only give the $3$-cells, since the $1$-cells and $2$-cells can be deduced from them. This polygraph is used in~\cite{BonfanteGuiraud06} to compute polynomials. The same document contains another example of polygraph, computing the fusion sort function on lists of natural numbers, which does not come from a TRS.
\end{example}

\subsection{Polygraphic interpretations}
\label{Subsection:Interpretations}

\noindent In order to prove that a polygraph terminates, we have developped a notion of polygraphic interpretation~\cite{Guiraud06jpaa}. Intuitively, we consider the $2$-paths as circuits crossed by currents. Each $2$-cell of a $2$-path produces some heat according to the intensity of the currents that reach it and the total heat produced by the generating $2$-cells of a $2$-path is used to compare it to other ones.

\begin{definition} \label{Definition:Interpretation}
Let $X$ and $Y$ be non-empty ordered sets and $M$ be a commutative monoid equipped with a strict, terminating order such that its addition is strictly monotone in both arguments. A \emph{polygraphic interpretation} of a polygraph $\Pr$ into $(X,Y,M)$ consists into a mapping of each $2$-path $f$ with $m$ inputs and $n$ outputs onto three monotone maps $f_* = \raisebox{-1.25mm}{\begin{picture}(0,0)%
\includegraphics{phi-bas.pstex}%
\end{picture}%
\setlength{\unitlength}{4144sp}%
\begingroup\makeatletter\ifx\SetFigFont\undefined%
\gdef\SetFigFont#1#2#3#4#5{%
  \reset@font\fontsize{#1}{#2pt}%
  \fontfamily{#3}\fontseries{#4}\fontshape{#5}%
  \selectfont}%
\fi\endgroup%
\begin{picture}(204,204)(1519,-2053)
\put(1621,-1985){\makebox(0,0)[b]{\smash{{\SetFigFont{6}{7.2}{\rmdefault}{\mddefault}{\updefault}{\color[rgb]{0,0,0}$f$}%
}}}}
\end{picture}%
}:X^m\fl X^n$, $f^* = \raisebox{-1.25mm}{\begin{picture}(0,0)%
\includegraphics{phi-haut.pstex}%
\end{picture}%
\setlength{\unitlength}{4144sp}%
\begingroup\makeatletter\ifx\SetFigFont\undefined%
\gdef\SetFigFont#1#2#3#4#5{%
  \reset@font\fontsize{#1}{#2pt}%
  \fontfamily{#3}\fontseries{#4}\fontshape{#5}%
  \selectfont}%
\fi\endgroup%
\begin{picture}(204,204)(1519,-2053)
\put(1621,-1985){\makebox(0,0)[b]{\smash{{\SetFigFont{6}{7.2}{\rmdefault}{\mddefault}{\updefault}{\color[rgb]{0,0,0}$f$}%
}}}}
\end{picture}%
}:Y^n\fl Y^m$ and $[f] = \raisebox{-1.25mm}{\begin{picture}(0,0)%
\includegraphics{phi-chaleur.pstex}%
\end{picture}%
\setlength{\unitlength}{4144sp}%
\begingroup\makeatletter\ifx\SetFigFont\undefined%
\gdef\SetFigFont#1#2#3#4#5{%
  \reset@font\fontsize{#1}{#2pt}%
  \fontfamily{#3}\fontseries{#4}\fontshape{#5}%
  \selectfont}%
\fi\endgroup%
\begin{picture}(204,204)(1519,-2053)
\put(1621,-1985){\makebox(0,0)[b]{\smash{{\SetFigFont{6}{7.2}{\rmdefault}{\mddefault}{\updefault}{\color[rgb]{0,0,0}$f$}%
}}}}
\end{picture}%
}:X^m\times Y^n\fl M$, such that the following conditions are satisfied:
\begin{itemize}
\item For every $1$-path $x$ of length $n$, we have $x_*=\id_{X^n}$, $x^*=\id_{Y^n}$ and $[x]=0$.
\item For every $2$-paths $f$ and $g$, the following three equalitities hold:
\begin{center}\input{definition-chaleurs-comp0.pstex_t}\end{center}

\item For every $2$-paths $f$ and $g$ such that $t_1(f)=s_1(g)$, the following three equalitities hold:
\begin{center}\input{definition-chaleurs-comp1.pstex_t}\end{center}

\item For every $3$-cell $\alpha:f\tfl g$, we have $f\succ g$, which means that, for every possible $x$ and $y$, the three inequalities $f_*(x) \geq g_*(x)$, $f^*(y) \geq g^*(y)$ and $[f](x,y) > [g](x,y)$ hold.
\end{itemize}
\end{definition}

\noindent The first three conditions in the definition of polynomial interpretation ensure that, given a $2$-path $f$, all the maps $f_*$, $f^*$ and $[f]$ are uniquely determined by the maps $\phi_*$, $\phi^*$ and $[\phi]$ for all the $2$-cells $\phi$ that $f$ is made of. The following result was proved for polygraphs with exactly one $1$-cell in~\cite{Guiraud06jpaa}. In~\cite{Guiraud06apal}, it was explained, on an example with two $1$-cells, how to extend the result to polygraphs with many $1$-cells.

\begin{theorem}[\cite{Guiraud06jpaa}] \label{Theorem:Interpretation}
If a $3$-polygraph $\Pr$ admits a polygraphic interpretation, then it terminates.
\end{theorem}

\begin{example} \label{Example:Interpretation}
Let us consider the polygraph consisting only of the four $3$-cells for addition and multiplication given in example~\ref{Example:Polygraph} and the following values:
\begin{itemize} 
\item $\figeps{cons-0}_*=1$, $\quad\figeps{cons-1}_*(i)=i+1$, $\quad\figeps{delta}_*(i)=(i,i)$, $\quad\figeps{fonction-2-b}_*(i,j)=i+j$,  $\quad\figeps{fonction-2}_*(i,j)=ij$;
\item $\left[\figeps{cons-0}\right] = \left[\figeps{cons-1}\right](i) = \left[\figeps{delta}\right](i) = \left[\figeps{epsilon}\right](i) = 0$, $\quad\left[\figeps{fonction-2-b}\right](i,j)=i$, $\quad\left[\figeps{fonction-2}\right](i,j)=(i+1)j$.
\end{itemize}

\noindent To prove that this yields a polygraphic interpretation into $(\Nb,\ast,\Nb)$, one makes computations such as:
$$
\left[ \raisebox{-2.75mm}{\includegraphics{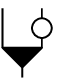}} \right] (i,j) \: = \: 
\left[ \figeps{fonction-2} \right] ( i, \figeps{cons-1}_*(j) ) + 
\left [\figeps{cons-1}\right](j) \: = \:
(i+1)(j+1).
$$

\noindent One can find more examples of polygraphic interpretations and related computations in~\cite{Guiraud06jpaa,BonfanteGuiraud06}. In~\cite{BonfanteGuiraud06}, it was proved that some polygraphic interpretations, such as the one we have built here, give more than termination: an information on the implicit complexity of the computed functions. Furthermore, the information we get here is divided into two parts: for every function $\figeps{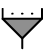}$, the current function $\figeps{phi}_*$ gives a bound on the size of the values computed by $\figeps{phi}$, while the heat function $\left[\figeps{phi}\right]$ limits the length of the computation. The heat function bound we get here is to be compared with the bounds that are found in~\cite{BonfanteCichonMarionTouzet01} by using usual polynomial interpretations on a TRS: $(i+1)j$ versus $(i+1)(j+1)$ for the multiplication and $i$ versus $2i+j+1$ for the addition.
\end{example}

\subsection{Polygraphic translations of term rewriting systems}
\label{Subsection:Translations}

This section recalls the standard translation of term rewriting systems into polygraphs. As a consequence of William Lawvere's work, term rewriting systems can be seen as presentations of algebraic theories~\cite{Lawvere63}. Then, Albert Burroni has proved that an algebraic theory can be presented by a $3$-polygraph~\cite{Burroni93}. Yves Lafont has given a standard translation of TRSs into $3$-polygraphs~\cite{Lafont03}. We have proved that this translation preserves termination and, under the hypothesis of left-linearity, reflects it~\cite{Guiraud06jpaa}. Because of size limitations, we only give here an informal construction of the polygraphic translation of a TRS, the formal results being in~\cite{Lafont03,Guiraud06jpaa}. 

Let us fix a (many-sorted) TRS $\Sigma=(\Sigma_1,\Sigma_2,\Sigma_3)$, with elementary sorts in $\Sigma_1$, operations in $\Sigma_2$ and rewriting rules in $\Sigma_3$. The \emph{standard polygraphic translation} of $\Sigma$ is denoted by~$\Pr(\Sigma)$ and is described thereafter, dimension after dimension:

Its $1$-cells are the elements of $\Sigma_1$.

Its $2$-cells are divided between \emph{algebra $\mathit{2}$-cells} and \emph{structure $\mathit{2}$-cells}. The algebra $2$-cells are the elements of $\Sigma_2$; if $\phi:\xi_1\times\cdots\times\xi_n\fl\xi$ is in $\Sigma_2$, then, as a $2$-cell, $\phi:\xi_1\star_0\cdots\star_0\xi_n\dfl\xi$. The family~$\Delta_2$ of \emph{structure $\mathit{2}$-cells} consists of one $\figeps{tau}:\xi\star_0\zeta\dfl\zeta\star_0\xi$ for each pair $(\xi,\zeta)$ of $1$-cells, plus one $\figeps{delta}:\xi\dfl\xi\star_0\xi$ and one $\figeps{epsilon}:\xi\dfl\ast$ for each $1$-cell $\xi$. Given a family $\vec{x}$ of distinct variables, any term~$u$ whose variables are in $\vec{x}$ admits a translation into a $2$-path $u_{\vec{x}}$. This is formalized in~\cite{Guiraud06jpaa} and suggested by the following examples, taken from the TRS for addition and multiplication of natural numbers:
$$
0_{\ast} \: = \: \figeps{cons-0}, \quad
s(x)_{xy} \: = \: \figeps{cons-1}\:\figeps{epsilon}, \quad
A(x,x)_x \: = \: \raisebox{-2.75mm}{\includegraphics{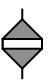}}, \quad
A(x,y)_{yx} \: = \: \raisebox{-2.75mm}{\includegraphics{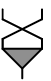}}, \quad
M(A(x,y),x)_{xyz} \: = \: \raisebox{-5.5mm}{\includegraphics{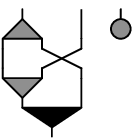}}.
$$

Its $3$-cells are divided between \emph{computation $\mathit{3}$-cells} and \emph{structure $\mathit{3}$-cells}. The computation $3$-cells are the elements of $\Sigma_3$; if $\alpha:u\fl v$ is in $\Sigma_3$ and $\vec{x}$ is the family of distinct variables appearing in $u$ from left to right, then $\alpha:u_{\vec{x}}\tfl v_{\vec{x}}$ when seen as a $3$-cell. The family $\Delta_3$ of structure $3$-cells is divided into two subfamilies. The first subfamily $\Delta_3^1$ depends only on the $1$-cells and is given by the following diagrams with each wire coloured by any possible $1$-cell:
\begin{center}\input{3-cellules-structure-1.pstex_t}\end{center}

\noindent The second subfamily $\Delta_3^2$ depends on the algebra $2$-cells and, for part, on the $1$-cells. It is given, for any algebra $2$-cell $\figeps{phi}$ and $1$-cell $\zeta$, by:
\begin{center}\input{3-cellules-structure-2.pstex_t}\end{center}

\noindent The $2$-targets of the $3$-cells of $\Delta_3^2$ use structure $2$-paths built from the structure $2$-cells by using the following structural induction rules:
\begin{center}\input{2-chemins-structure.pstex_t}\end{center}

\begin{proposition}[\cite{Guiraud06jpaa}] \label{Proposition:TranslationProperties}
Let $\Sigma$ be a term rewriting system. Then the following properties hold:
\begin{enumerate}
\item The three families $\Delta_3^1$, $\Delta_3^2$ and $\Delta_3$ are convergent.

\item For any term $u$ and any possible family $\vec{x}$ of variables, $u_{\vec{x}}$ is a $\Delta_3$-normal form. 

\item Any $\Delta_3^2$ normal form is of the shape $f\star_1 g$ with $f$ in $\mon{\Delta_2}$ and $g$ in $\mon{\Sigma_2}$.

\item Any $\Delta_3$-normal form is of the shape $f\star_1 g$ with $f$ a $\Delta_3^1$-normal form in $\mon{\Delta_2}$ and $g$ in $\mon{\Sigma_2}$.

\item If $\Sigma$ is left-linear, $u$ and $v$ are two terms and $\alpha$ is a rule such that $u\fl_{\alpha}v$ holds, then, for any possible family $\vec{x}$ of variables, there exists a $2$-path $f$ such that $u_{\vec{x}} \: \tfl_{\alpha}^1 \: f \: \tfl_{\Delta_3} \: v_{\vec{x}}$ holds.
\end{enumerate}

\end{proposition}

\noindent In the following equivalence, the direct implication is true even without the hypothesis of left-linearity: it is proved by using a special polygraphic interpretation. The reverse direction is the one that is of interest for us in this study and is proved with the help of proposition~\ref{Proposition:TranslationProperties} last point.

\begin{theorem}[\cite{Guiraud06jpaa}] \label{Theorem:Termination}
A left-linear TRS $\Sigma$ terminates if and only $\Pr(\Sigma)$ does.
\end{theorem}

\begin{example} \label{Example:InterpretationTry}
Let us try to use theorem~\ref{Theorem:Termination} to prove termination of the division program and see why and how we want to enhance it. We consider the polygraphic interpretation of the computation $3$-cells into $(\Nb,\ast,\Nb)$ generated by the following values:
\begin{itemize}
\item $\figeps{cons-0}_*=1$, $\:\figeps{cons-1}_*(i)=(i+2)$, $\:\figeps{tau}_*(i,j)=(j,i)$, $\:\figeps{delta}_*(i)=(i,i)$, $\:\figeps{fonction-2-b}_*(i,j)=\figeps{fonction-2}_*(i,j)=i$;
\item $\left[\figeps{cons-0}\right]=\left[\figeps{cons-1}\right](i)=\left[\figeps{tau}\right](i,j)=\left[\figeps{delta}\right](i)=\left[\figeps{epsilon}\right](i)=0$, $\:\left[\figeps{fonction-2-b}\right](i,j)=j$, $\:\left[\figeps{fonction-2}\right](i,j)=ij$.
\end{itemize}

\noindent One can check that this yields an interpretation such that $s_2(\alpha)\succ t_2(\alpha)$ for any computation $3$-cell $\alpha$. For example, for the last computation $3$-cell $\alpha$, we have both  $s_2(\alpha)_* (i,j)$  and  $t_2(\alpha)_*(i,j)$ equal to $i+2$ and $[s_2(\alpha)](i,j)=ij+2j$ while $[t_2(\alpha)](i,j)=ij+j$. But the $3$-cells expressing how to duplicate $\figeps{fonction-2-b}$ and $\figeps{fonction-2}$ satisfy the reverse strict inequality: we do not have a polygraphic interpretation of the standard polygraphic translation. 

Thus, we do not have its termination and, consequently, no information on the implicit complexity of the division function, even if the interpretation we have considered satisfies the conditions of~\cite{BonfanteGuiraud06}. In section~\ref{Section:Results}, we correct this problem with results that allow us, in particular, to conclude for the present example.
\end{example}

\section{Reduced polygraphic translations for left-linear term rewriting systems} 
\label{Section:Results}

From now on, we assume that $\Sigma$ is a left-linear term rewriting system.

\subsection{The general case}
\label{Subsection:GeneralCase}

Here we prove that there is no need to consider the family $\Delta_3^1$ of structure $3$-cells for proving termination of the original TRS: these $3$-cells are only required to ensure confluence. 

\begin{lemma}\label{Lemma:FactorisationStructure}
Let $f$ be a $2$-path and $g$ its $\Delta_3$-normal form. Then there exists a $2$-path $h$ in $\Delta_3^2$-normal form such that $f \: \tfl_{\Delta_3^2} \: h \: \tfl_{\Delta_3^1} \: g$ holds.
\end{lemma}

\begin{proof}
Let $h$ be the $\Delta_3^2$-normal form of $f$. We use proposition~\ref{Proposition:TranslationProperties}: since $\Delta_3$ is confluent (first point), we know that the $\Delta_3$-normal form of $h$ is $g$. From the shape of $h$ (third point), we deduce that only $3$-cells from $\Delta_3^1$ can be applied to $h$ or any of its reduces.
\end{proof}

\begin{theorem}\label{Theorem:GeneralCase}
If $\Pr(\Sigma)$ terminates without the first family of structure $3$-cells, then $\Sigma$ terminates.
\end{theorem}

\begin{proof}
Let us assume that $\Pr(\Sigma)$ terminates without $\Delta_3^1$ but that $\Sigma$ does not terminate. Then, there exists a sequence $(u_n)_{n\in\Nb}$ of terms and a sequence $(\alpha_n)_{n\in\Nb}$ such that, for every natural number $n$, we have $u_n\fl_{\alpha_n}u_{n+1}$. Let us fix a family $\vec{x}$ of variables such that $(u_0)_{\vec{x}}$ is defined. Then, the last point of proposition~\ref{Proposition:TranslationProperties} yields a family of $2$-paths $(f_n)_{n\in\Nb}$ such that $(u_n)_{\vec{x}} \: \tfl_{\alpha_n}^1 \: f_n \: \tfl_{\Delta_3} \: (u_{n+1})_{\vec{x}}$ holds for every~$n$. Let us fix a natural number~$n$. From proposition~\ref{Proposition:TranslationProperties}, we know that $(u_{n+1})_{\vec{x}}$ is a $\Delta_3$-normal form (second point) and that $\Delta_3$ is confluent (first point): hence $(u_{n+1})_{\vec{x}}$ is the $\Delta_3$-normal form of the $2$-path $f_n$. We apply lemma~\ref{Lemma:FactorisationStructure} and get a $2$-path $g_n$ in $\Delta_3^2$-normal form satisfying:
$$
(u_n)_{\vec{x}} \: \tfl_{\alpha_n}^1 \: f_n \: \tfl_{\Delta_3} \: g_n \: \tfl_{\Delta_3^1} \: (u_{n+1})_{\vec{x}}.
$$

\noindent With proposition~\ref{Proposition:TranslationProperties}, we know that $g_n=h_n\star_1k_n$ (third point) and $(u_{n+1})_{\vec{x}}=h'_n\star_1 k_n$ (fourth point), with $k_n$ in $\mon{\Sigma_2}$, $h_n$ in $\mon{\Delta_2}$ and $h'_n$ its $\Delta_3^1$-normal form. But we have seen that $(u_{n+1})_{\vec{x}}\tfl_{\alpha_{n+1}}^1 f_{n+1}$ holds and, since the TRS we consider is left-linear, the $2$-source of $\alpha_{n+1}$ does not contain any structure $2$-cell. This implies that $s_2(\alpha_{n+1})$ is entirely contained into $k_n$, so that $f_{n+1}$ can be decomposed into $h'_n\star_1 k'_n$, with $k_n\tfl_{\alpha_{n+1}}^1 k'_n$. We deduce from these facts the following reduction chain:
$$
g_n=h_n\star_1 k_n \: \tfl_{\alpha_{n+1}}^1 \: h_n\star_1 k'_n \: \tfl_{\Delta_3^1} \: h'_n\star_1 k'n = f_{n+1}.
$$

\noindent We know that $(u_{n+2})_{\vec{x}}$ is the $\Delta_3$-normal form of $f_{n+1}$. By confluence of $\Delta_3$, we deduce that it is also the $\Delta_3$-normal form of $h_n\star_1 k'_n$. Then lemma~\ref{Lemma:FactorisationStructure} gives the existence of a $\Delta_3^2$-normal form $g_{n+1}$ such that:
$$
h_n\star_1 k'_n \: \tfl_{\Delta_3^2} \: g_{n+1} \: \tfl_{\Delta_3^1} \: (u_{n+2})_{\vec{x}}.
$$

\noindent By induction on $n$, we conclude that the infinite reduction path $(u_n)_{n\in\Nb}$ in $\Sigma$ generates an infinite reduction path in $\Pr(\Sigma)$ that only uses $3$-cells of $\Sigma_3$ and $\Delta_3^2$, the existence of which has been prohibited by assumption. Hence $\Sigma$ terminates.
\end{proof}

\begin{remark}
Removing some structure $3$-cells allows for a wider range of interpretations for the duplication, such as $\figeps{delta}_*(i)=\left(\roundup{i/2},\rounddown{i/2}\right)$ for the descending currents. This would be prohibited by the two structure $3$-cells of $\Delta_3^1$ involving both $\figeps{delta}$ and $\figeps{epsilon}$, since they require the inequality $\figeps{delta}_*(i)\geq(i,i)$.
\end{remark}

\subsection{The case of planar linear term rewriting systems}
\label{Subsection:PlanarLinearCase}

Let us recall that a term rewriting rule is usually called linear when no variable occur twice in its left member or in its right member. Here and in order to match the vocabulary from linear algebra and operadic theory, we call a rule \emph{linear} when its two sides contain exactly the same variables, exactly once. Thus, if $\Sigma$ is linear, which means that all of its rules are, the computation $3$-cells of the polygraph $\Pr(\Sigma)$ do not use any $\figeps{delta}$ or~$\figeps{epsilon}$. We say that a term rewriting rule is \emph{planar} when variables occur in the same order in its two sides. Then $\Sigma$ is planar when all of its rules are and, in that case, the computation $3$-cells of $\Pr(\Sigma)$ do not use any~$\figeps{tau}$.

\begin{theorem}\label{Theorem:PlanarLinearCase}
Let us assume that $\Sigma$ is both linear and planar. If $\Pr(\Sigma)$ terminates without the structure $3$-cells, then $\Sigma$ terminates.
\end{theorem}

\begin{proof}
If $\Sigma$ is both linear and planar, then the computation $3$-cells of $\Pr(\Sigma)$ do not contain any structure $2$-cell. Now, let us assume that $\Pr(\Sigma)$ terminates without the structure $3$-cells. If $\Sigma$ does not terminate, then there exists an infinite reduction path $u_0 \fl_{\alpha_0} u_1 \fl_{\alpha_1} u_2 \fl_{\alpha_2} (\cdots).$ Let us fix a family $\vec{x}$ of distinct variables such that $(u_0)_{\vec{x}}$ is defined. By proposition~\ref{Proposition:TranslationProperties}, points two and four, each $(u_n)_{\vec{x}}$ decomposes into $g_n\star_1 h_n$ with $g_n$ in $\mon{\Delta_2}$ and $h_n$ in $\mon{\Sigma_2}$. Let us fix a natural number $n$. Then, point five of the same proposition yields a $2$-path $f_n$ such that:
$$
g_n\star_1 h_n = (u_n)_{\vec{x}} \: \tfl_{\alpha_n}^1 \: f_n \: \tfl_{\Delta_3} \: (u_{n+1})_{\vec{x}} = g_{n+1}\star_1 h_{n+1}.
$$

\noindent Since $\alpha_n$ is left-linear, its $2$-source does not contain any structure $2$-cell, so that it is entirely contained into $h_n$. This means that there exists a $2$-path $h'_n$ such that $h_n\tfl_{\alpha_n}^1 h'_n$ and $f_n=g_n\star_1 h'_n$. Since $\alpha_n$ is linear and planar, its $2$-target does not contain any structure $2$-cell either: this means that $h'_n$ is in $\mon{\Sigma_2}$. Hence, any $3$-path of $\mon{\Delta_3}$ starting at $f_n=g_n\star_1 h'_n$ only acts on $g_n$. So $g_{n+1}$ is the $\Delta_3$-normal form of $g_n$ and $h'_n=h_{n+1}$. Thus, we have $h_n\tfl_{\alpha_n}^1 h_{n+1}$. Since this is valid for any natural number $n$, we have an infinite reduction path in $\Pr(\Sigma)$ that does not use any structure $3$-cell: this is prohibited by our hypothesis, so that $\Sigma$ terminates.
\end{proof}

\begin{remark}\label{Remark:PlanarLinearCase}
The proof of theorem~\ref{Theorem:PlanarLinearCase} can be adaptated for TRSs whose computation $3$-cell do not require any~$\figeps{tau}$ or any~$\figeps{delta}$ or any~$\figeps{epsilon}$. In that case, we do not discard all the srtucture $3$-cells of $\Delta_3^2$ but only the ones concerning the unused structure $2$-cell(s).
\end{remark}

\begin{example}
The TRS for the double function is both linear, in our sense, and planar. Hence, for proving its termination, we only have to prove the termination of the two computation $3$-cells of its translation. We consider the interpretation into $(\Nb,\ast,\Nb)$ generated by $\figeps{cons-0}_*=1$, $\figeps{cons-1}_*(i)=i+1$, $\figeps{fonction-1}_*(i)=2i$, $\left[\figeps{cons-0}\right]=0$, $\left[\figeps{cons-1}\right](i)=0$, $\left[\figeps{fonction-1}\right](i)=i$. On top of proving termination, this polygraphic interpretation satisfies the conditions given in~\cite{BonfanteGuiraud06} that allow us to conclude that the polynomials $\figeps{fonction-1}_*(i)=2i$ and $\left[\figeps{fonction-1}\right](i)=i$ respectively bound the size of the computed values and the length of the computations, with respect to the size of the argument. This is to be compared with the polynomial bounds of $3i$ given by polynomial interpretations on terms and of $2i$ given by the same interpretations with a preprocessing using dependency pairs.
\end{example}

\subsection{The case of first-order functional programs}
\label{Subsection:FunctionalProgramCase}

\begin{definition}
A \emph{function} (or \emph{defined symbol}) of $\Sigma$ is an operation in $\Sigma_2$ that only appears as the root symbol of the left-hand side of rewriting rules of $\Sigma_3$. A \emph{constructor} of $\Sigma$ is an operation in $\Sigma_2$ that never appears as the root symbol of the left-hand side of rewriting rules of $\Sigma_3$. Two rules $\alpha$ and $\beta$ are \emph{weakly orthogonal} when all their critical pairs are of the form $u\rightrightarrows^{\alpha}_{\beta} v$: whenever one can apply $\alpha$ and $\beta$ on overlapping parts of the same term, both reductions give the same result. 

A \emph{first-order functional program} is a left-linear term rewriting system $\Sigma$ whose operations are either a function or a constructor and whose rewriting rules are pairwise weakly orthogonal. In that case, we denote by $\Sigma_2^C$ the set of constructors and by $\Sigma_2^F$ the one of functions. The \emph{polygraphic program} associated to a first-order functional program $\Sigma$ is the standard polygraphic translation $\Pr(\Sigma)$ without $\Delta_3^1$ and $\Delta_3^F$, the $3$-cells of $\Delta_3^2$ corresponding to functions. We denote by $\Delta_3^C$ the $3$-cells of $\Delta_3^2$ corresponding to constructors.
\end{definition}

\noindent The notion of polygraphic program was introduced in~\cite{BonfanteGuiraud06}. This class of rewriting systems contains more than translations of first-order functional programs: polygraphic programs can compute functions with many outputs with a link between them, such as the list splitting function that is studied in~\cite{BonfanteGuiraud06}. A key argument for the following result is given in~\cite{Gramlich94}: for a first-order functional program, termination and innermost termination are equivalent.

\begin{theorem}\label{Theorem:FunctionalProgramCase}
Let $\Sigma$ be a first-order functional program. If the polygraphic program associated to $\Sigma$ terminates, then so does $\Sigma$.
\end{theorem}

\begin{proof}
Let us assume that the polygraphic program associated to $\Sigma$ terminates but that $\Sigma$ does not. Hence, there exists an infinite innermost reduction sequence $u_0 \fl_{\alpha_0}^i u_1 \fl_{\alpha_1}^i  u_2  \fl_{\alpha_2}^i (\cdots{})$ in $\Sigma$. Let $\vec{x}$ be a family of variables such that $(u_0)_{\vec{x}}$ is defined. Point five of proposition~\ref{Proposition:TranslationProperties} tells us that the reduction sequence $(u_n)_{n\in\Nb}$ lifts up to $\Pr(\Sigma)$, yielding $(u_0)_{\vec{x}} \: \tfl_{\alpha_0}^1 \: f_0 \:\tfl_{\Delta_3} \: (u_1)_{\vec{x}} \: \tfl_{\alpha_1}^1 \: f_1 \:\tfl_{\Delta_3}  \: (u_2)_{\vec{x}} \: \tfl_{\alpha_2}^1 (\cdots{})$. Since $\Sigma$ is a functional program, there exist, for every $n$ in $\Nb$, $\phi_n$ in $\Sigma_2^F$, $a_n$ in $\mon{\Sigma_2^C}$, $a'_n$ in $\mon{\Sigma_2}$ and~$b_n$ in $\mon{\Delta_2}$ such that $s_2(\alpha_n)=a_n\star_1\phi_n$ and $t_2(\alpha_n)=b_n\star_1 a'_n$. Now, since $(u_n)_{\vec{x}}\tfl_{\alpha_n}^1 f_n$, there exist~$C_n$ and $c_n$ in $\mon{\Sigma_2}$ and $S_n$ in $\mon{\Delta_2}$ such that:
\begin{center}\input{decomposition-1.pstex_t}\end{center}

\noindent By examination of the shapes and properties of the structure $3$-cells, there exist $b'_n$, $b''_n$ and $S'_n$ in $\mon{\Delta_2}$,~$c'_n$ in $\mon{\Sigma_2^C}$, $d_n$ and $d'_n$ in $\mon{\Sigma_2}$ such that the following three \emph{normalizing} reductions hold:
\begin{center}\input{decomposition-2.pstex_t}\end{center}

\noindent Thus, we have two decompositions of $(u_{n+1})_{\vec{x}}$, expressed by the following equalities:
\begin{center}\input{decomposition-3.pstex_t}\end{center}

\noindent In the leftmost decomposition, the function $2$-cell $\phi_{n+1}$ of the rightmost decomposition must appear in either $d'_n$, $a'_n$ or $C_n$, since $c'_n$ is in $\Sigma_2^C$ and $S'_n$ is in $\mon{\Delta_2}$. Let us assume that it is in $d'_n$. Since $d'_n$ has been produced from $d_n$ by the action of the structure $2$-cells, any redex it contains is a copy of one that is already in $d_n$. This means that the reduction $\alpha_{n+1}$ can already be applied on $u_n$, in a proper subterm of the term where $\alpha_n$ is applied: this is in contradiction with the hypothesis that the reduction from $u_n$ to $u_{n+1}$ is innermost. Hence, $\phi_{n+1}$ is in $C_n$ or in $a'_n$. Furthermore, $a_{n+1}$ is only made of constructors: each $2$-cell it contains was either already in $C_n$ or $a'_n$ or appear in $c'_n$ as the result of the application of $3$-cells of $\Delta_3^C$. Hence, the reduction $(u_{n+1})_{\vec{x}}\tfl_{\alpha_{n+1}}^1 f_{n+1}$ can be anticipated on the following $2$-path~$g_n$, which is the $\Delta_3^C$-normal form of $f_n$:
\begin{center}\input{decomposition-4.pstex_t}\end{center}

\noindent Let us denote by $h_n$ the result of this $\alpha_{n+1}$-reduction on $g_n$ and by $g_{n+1}$ the $\Delta_3^C$-normal form of $h_n$. Then $g_{n+1}$ can be normalized successively by $\Delta_3^2$ and by $\Delta_3^1$ to reach $(u_{n+2})_{\vec{x}}$. Using again the fact that the reductions on terms have been supposed to be innermost, we prove that the reduction acting on $(u_{n+2})_{\vec{x}}$ can also be anticipated on $g_{n+1}$, yielding $h_{n+1}$ and so on. Thus, an induction on $n$ gives an infinite reduction sequence $(u_0)_{\vec{x}} \: \tfl_{\alpha_0}^1 \: f_0 \: \tfl_{\Delta_3^C} \: g_0 \: \tfl_{\alpha_1}^1 \: h_0 \: \tfl_{\Delta_3^C} \: g_1 \: \tfl_{\alpha_2}^1 \: h_1 \: \tfl_{\Delta_3^C} \: (\cdots)$, which cannot exist by termination of the polygraphic program associated to $\Sigma$.
\end{proof}

\begin{example}
Let us consider the interpretation we have built in example~\ref{Example:InterpretationTry}. Now, equipped with theorem~\ref{Theorem:FunctionalProgramCase}, we can conclude that this interpretation proves the termination of the original term rewriting system for division. Moreover, using the results proved in~\cite{BonfanteGuiraud06}, we conclude that the polynomials $\figeps{fonction-2}_*(X,Y)=X$ and $\left[\figeps{fonction-2}\right](X,Y)=XY$ respectively bound the spatial and temporal sizes of the computation of the division on two arguments with sizes $X$ and $Y$.
\end{example}

\begin{remark}
The result is false if one removes the weak orthogonality assumption. Indeed, let us consider the polygraphic program whose computation $3$-cells are:
\begin{center}\input{contre-exemple.pstex_t}\end{center}

\noindent We prove that it terminates with a mapping into $(\Nb,\ast,[\Nb])$, where $[\Nb]$ is the free commutative monoid generated by $\Nb$ with its natural multiset order, with $\ens{n}$ standing for $n$ seen as a generator of $[\Nb]$. We consider: $\figeps{fonction-1}_*(i)=i$, $\figeps{cons-1}_*(i) = \figeps{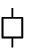}_*(i) = \figeps{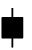}_*(i) = i+1$, $\figeps{fonction-2}_*(i,j)=i+j$, $\figeps{delta}_*(i)=(\roundup{i/2},\rounddown{i/2})$ and  $\left[\figeps{cons-1}\right](i) = \left[\figeps{cons-1-b}\right](i) = \left[\figeps{delta}\right](i) = \left[\figeps{epsilon}\right](i) = 0$, $\left[\figeps{fonction-1}\right](i) = \left[\figeps{fonction-1-b}\right](i) = \ens{i}$ and $\left[\figeps{fonction-2}\right](i,j)=\ens{i+j}$. One proves that this mapping satisfies $s_2(\alpha)\succ t_2(\alpha)$ if $\alpha$ is the second, third or fourth computation $3$-cell and that $s_2(\beta)=t_2(\beta)$ is $\beta$ is the first one or is in $\Delta_3^C$. Then, we define the mapping counting the number of $\figeps{fonction-1}$ in a $2$-path: this is the mapping into $(\ast,\ast,\Nb)$ whose heat function sends any $2$-cell to~$0$ except $\figeps{fonction-1}$, sent to $1$. Finally, we use the termination of $\Delta_3^C$ to get the one of the polygraphic program. However, if we add $\Delta_3^F$ and $\Delta_3^1$, it does not terminate anymore, as proved by the following cycle:
\begin{center}\input{cycle.pstex_t}\end{center}

\noindent However, the results in~\cite{Gramlich94} seem to indicate that the weak orthogonality hypothesis is too strong and could be replaced by local confluence.
\end{remark}

\subsection{Special conditions on standard interpretations}
\label{Subsection:SpecialInterpretations}

Until now, we have seen conditions based solely on the properties of the original TRS. Here we assume that we use polygraphic interpretations and give conditions on them: the purpose of such results is to guide the automatic search of polygraphic interpretations. Let us recall that a TRS is \emph{non-duplicating} when no right-hand side of a rewriting rule contains the same variable twice. Hence, the computation $3$-cells of such a TRS polygraphic translation do not use any~$\figeps{delta}$.

\begin{theorem} \label{Theorem:NonDuplicating}
Let us assume that $\Sigma$ is non-duplicating. If $\Pr(\Sigma)$, without the structure $3$-cells, admits a polygraphic interpretation into some $(X,\ast,M)$ such that $\figeps{tau}_*(x,y)=(y,x)$, then $\Sigma$ terminates.
\end{theorem}

\begin{proof}
Since $\Sigma$ is non-duplicating, we adapt the proof of~\ref{Theorem:PlanarLinearCase}, as mentionned in remark~\ref{Remark:PlanarLinearCase}, to get that the termination of $\Sigma$ can be deduced from the one of $\Pr(\Sigma)$, without $\Delta_3^1$ and all the structure $3$-cells of~$\Delta_3^2$ that concern $\figeps{delta}$; for this proof, we denote by $\Delta_3^n$ the remaining structure $3$-cells. Now, let us assume that $\Pr(\Sigma)$, without the structure $3$-cells, admits a polygraphic interpretation $((\cdot)_*,(\cdot)^*,[\cdot])$ into $(X,\ast,M)$ such that $\figeps{tau}_*(i,j)=(j,i)$ holds. Then we define another mapping into $(X,\ast,M)$ with the same currents functions and with a heat function $\ens{\cdot}$ defined as $[\cdot]$ except on the structure $2$-cells, which it sends to $0$. Since the addition of $M$ is monotone in each argument and by properties of the heat functions, we have $[f]\geq\ens{f}$ for any $2$-path $f$, with equality when $f$ is in $\mon{\Sigma_2}$. Hence, by left-linearity of $\Sigma$, we get, for each computation $3$-cell $\alpha$, $\ens{s_2(\alpha)} = [s_2(\alpha)] > [t_2(\alpha)] \geq \ens{t_2(\alpha)}$. Now let us fix an algebra $2$-cell $\figeps{phi}$ and consider the structure $3$-cells of $\Delta_3^n$ it is involved into. For $\figeps{epsilon}$, the current functions are equal on both sides and $\left\{\figeps{phi}\star_1\figeps{epsilon}\right\}(\vec{x}) = \left[\figeps{phi}\right](\vec{x}) \geq 0  = \left\{\figeps{epsilon}\star_0\cdots\star_0\figeps{epsilon}\right\} (\vec{x})$. For $\figeps{tau}$, we have equality between the current functions and the heat functions on both sides, thanks to the hypothesis $\figeps{tau}_*(x,y)=(y,x)$. Hence, the new map $\ens{\cdot}$ generates a terminating order relation $\succ$ on $\Pr(\Sigma)$ such that, for every computation $3$-cell, $s_2(\alpha)\succ t_2(\alpha)$ and, for every structure $3$-cell $\beta$ of $\Delta_3^n$, $s_2(\beta)\succeq t_2(\beta)$. Thus, all these $3$-cells, together, terminate if and only if $\Delta_3^n$ terminates, which is true.
\end{proof}

\begin{notation}
Let $\alpha:\phi(u_1,\dots,u_n)\fl v$ be a rewrite rule in $\Sigma_3$. For $i\in\ens{1,\dots,n}$ we denote by~$K_i(\alpha)$ the greatest of the number of occurences in $v$ of each variable of $u_i$. For any function $\phi$ of arity $n$ in $\Sigma_2$ and any $i\in\ens{1,\dots,n}$, we denote by $K_i(\phi)$ the greatest of the $K_i(\alpha)$ for all the rules $\alpha$ such that~$\phi$ is the root symbol of the left-hand side of $\alpha$. 
\end{notation}

\begin{theorem}\label{Theorem:SpecialInterpretations}
Let $X$ be a set equipped with a terminating strict order. Let us assume that $\Pr(\Sigma)$, without the structure $3$-cells, admits a polygraphic interpretation into $(X,\ast,[X])$ such that the following conditions hold:
\begin{itemize}
\item $\figeps{delta}(x)=(x,x)$, $\figeps{tau}(x,y)=(y,x)$ and $[\sigma]=0$ when $\sigma\in\left\{\figeps{tau},\figeps{delta},\figeps{epsilon}\right\}$.
\item $\figeps{phi}_*(x_1,\dots,x_n) \geq x_i$ for all $\figeps{phi}$ and all $i$.
\item $\left\{\figeps{phi}_*(\vec{x})\right\} > \left[\figeps{phi}\right](\vec{x})$ for all $\figeps{phi}$.
\item $\left[\figeps{phi}\right](x_1,\dots,x_n) > \ens{x_i}$ if $\phi$ is the root symbol of some rule and $K_i(\phi)\geq 2$.
\end{itemize}

\noindent Then $\Sigma$ terminates.
\end{theorem}

\begin{proof}
Let $a$ and $b$ be two terms and $\alpha$ a rewriting rule such that $a\fl_{\alpha}b$. We denote by $u=\phi(u_1,\dots,u_p)$ the source of $\alpha$, by $v$ its target and by $\vec{x}$ and $\vec{y}$ the respective families of distinct variables that appear in $u$ and $v$ from left to right. We fix a family $\vec{z}$ of distinct variables containing all the variables that appear in~$a$. Then the $2$-paths $a_{\vec{z}}$ and $b_{\vec{z}}$ decompose as:
\begin{center}\input{decomposition-5.pstex_t}\end{center}

\noindent In these decompositions, $S$ and $S'$ are in $\mon{\Delta_2}$ and $f$, $g$, $C$ and $C'$ in $\mon{\Sigma_2}$. Let us write $f=f_1\star_0\cdots\star_0 f_m$ and $g=g_1\star_0\cdots\star_0 g_n$, where each $f_i$ and $g_j$ has a $1$-cell as $1$-target. We denote by $j(i)$ the element of $\ens{1,\dots,p}$ such that $f_i$ appears inside $u_{j(i)}$ in $a$. Let $c$ be the structure $2$-path and $d$ the algebra $2$-path such that $v_{\vec{x}}=c\star_1 d$, given by proposition~\ref{Proposition:TranslationProperties} points two and four. Using the same proposition, we know that there exists a structure $2$-path $c'$ such that $c'\star_1 g$ is the $\Delta_3^2$-normal form of $f\star_1 c$. But the structure $3$-cells of $\Delta_3^2$ act in such a way that each $g_j$ is exactly one $f_i$. Moreover, in the family $(g_1,\dots,g_n)$, each $f_i$ appears at most $K_{j(i)}(\phi)$ times, by definition of $K_{j(i)}(\phi)$. 

Now, let us compute the interpretations of $a_{\vec{y}}$ and $b_{\vec{y}}$. By hypothesis on the interpretation, we have $(a_{\vec{y}})_*\geq (b_{\vec{y}})_*$. Concerning the heats, still using the assumptions on the interpretation, $C$ receives at least the same currents in $a_{\vec{y}}$ than in $b_{\vec{y}}$ and $[C]$ is monotone: hence $C$ produces at least the same heat in~$a_{\vec{y}}$ than in $b_{\vec{y}}$. For the same reasons and since $[s_2(\alpha)]>[t_2(\alpha)]$, $u_{\vec{x}}$ produces strictly more heat in~$a_{\vec{y}}$ than in $b_{\vec{y}}$. Furthermore $f$ produces $\sum_{i=1}^m [f_i](\vec{k}_i)$ while $g$ produces $\sum_{i=1}^m K_{j(i)}(\phi).[f_i](\vec{k}_i)$ for some $\vec{k_i}$: indeed the currents received by each copy of $f_i$ in $g$ are the same as the currents received by the original~$f_i$, by properties of the current map on structure $2$-cells and by examination of each structure $3$-cell. Finally, the structure $2$-paths $S$ and $S'$ do not produce any heat.

Since we consider a multiset order on $[X]$, we can prove that $[a_{\vec{y}}]>[b_{\vec{y}}]$ by proving that $u_{\vec{y}}$ produces strictly more heat than each $f_i$ such that $K_{j(i)}(\phi)\geq 2$; by property of the heat function, it is even sufficient to prove that $[\phi]$ produces more heat than each $f_i$ such that $K_{j(i)}\geq 2$. In $a_{\vec{y}}$, the current received by~$u_{\vec{y}}$ is $((f_1)_*(\vec{k}_1),\dots,(f_m)_*(\vec{k})_m)$. Since each $u_j$ transmits a current at least equal to the one it receives in any of its inputs, $\phi$ receives at least $(f_i)_*(\vec{k}_i)$ in its imput $j(i)$. By assumption, if $K_{j(i)}(\phi)\geq 2$, then the heat produced by $\phi$ is strictly greater than $\ens{(f_i)_*(\vec{k}_i)}$, which, in turn, is striclty greater than $[f_i](\vec{k}_i)$, once again since we consider a multiset order on $[X]$. Finally, we have $a_{\vec{y}}\succ b_{\vec{y}}$ whenever $a\fl_{\alpha}b$: we deduce from this fact that $\Sigma$ terminates.
\end{proof}

\section{Conclusion}
\label{Section:Conclusion}

In this study, we have proved results that make easier the use of polygraphs and polygraphic interpretations for proving termination for TRSs and, when it comes to functional programs, finding an implicit complexity bound with some interpretations. We have seen, on some examples, that the method can give better results than polynomial interpretations on TRSs, mainly for functional programs: it gives better complexity bounds and can prove the termination and give bounds for TRSs that do not admit simplification orders.

The next step consists into a test of this method on the Termination Problems DataBase in order to get information on its efficiency and to formulate new conjectures. Among them, we want to examine the hypothesis of weak orthogonality used in theorem~\ref{Theorem:FunctionalProgramCase} and new ways to guide the construction of the polygraphic interpretation with respect to the shape of the rewriting rules, in the same spirit as in~\ref{Subsection:SpecialInterpretations}. We also plan to enhance the theoretical links between termination of TRSs and termination of polygraphic versions of them: indeed, we think that there are many known, finer results on term graph rewriting systems that can be adaptated to polygraphs~\cite{Plump99}. The case of non left-linear TRSs may be examinated but we are not sure that the polygraphic translations can provide methods for them.

Finally, we need a better understanding of the mathematical structure behind the one of polygraphic interpretation. This will allow for new kinds of interpretations, extending the range of the method for a wider variety of functional programs. The main class we will focus on are ones with conditional rules or with the if-then-else construction, in order to solve some of the problems we have encountered in~\cite{BonfanteGuiraud06}.

\bibliographystyle{amsplain}
\bibliography{bibliographie}

\providecommand{\bysame}{\leavevmode\hbox to3em{\hrulefill}\thinspace}
\providecommand{\MR}{\relax\ifhmode\unskip\space\fi MR }
\providecommand{\MRhref}[2]{%
  \href{http://www.ams.org/mathscinet-getitem?mr=#1}{#2}
}
\providecommand{\href}[2]{#2}
\begin{thebibliography}{10}

\bibitem{ArtsGiesl00}
Thomas Arts and Jürgen Giesl, \emph{Termination of term rewriting using
  dependency pairs}, Theoretical Computer Science \textbf{236} (2000), no.~1-2,
  133--178.

\bibitem{BaaderNipkow98}
Franz Baader and Tobias Nipkow, \emph{Term rewriting and all that}, Cambridge
  University Press, 1998.

\bibitem{BonfanteCichonMarionTouzet01}
Guillaume Bonfante, Adam Cichon, Jean-Yves Marion, and Hélène Touzet,
  \emph{Algorithms with polynomial interpretation termination proofs}, Journal
  of Functional Programming \textbf{11} (2001), no.~1, 33--53.

\bibitem{BonfanteGuiraud06}
Guillaume Bonfante and Yves Guiraud, \emph{Programs as polygraphs:
  computability and complexity}, Submitted, 2006.

\bibitem{Burroni93}
Albert Burroni, \emph{Higher-dimensional word problems with applications to
  equational logic}, Theoretical Computer Science \textbf{115} (1993), no.~1,
  43--62.

\bibitem{CichonLescanne92}
Adam Cichon and Pierre Lescanne, \emph{Polynomial interpretations and the
  complexity of algorithms}, Lecture Notes in Artificial Intelligence
  \textbf{607} (1992), 139--147.

\bibitem{Gramlich94}
Bernhard Gramlich, \emph{On modularity of termination and confluence properties
  of conditional rewrite systems}, Lecture Notes in Computer Science
  \textbf{859} (1994), 186--203.

\bibitem{Guiraud06jpaa}
Yves Guiraud, \emph{Termination orders for 3-dimensional rewriting}, Journal of
  Pure and Applied Algebra \textbf{207} (2006), no.~2, 341--371.

\bibitem{Guiraud06apal}
\bysame, \emph{The three dimensions of proofs}, Annals of Pure and Applied
  Logic \textbf{141} (2006), no.~1-2, 266--295.

\bibitem{Guiraud06tcs}
\bysame, \emph{Two polygraphic presentations of petri nets}, Theoretical
  Computer Science \textbf{360} (2006), no.~1-3, 124--146.

\bibitem{Lafont90}
Yves Lafont, \emph{Interaction nets}, Principles of Programming Languages, ACM
  Press, 1990, pp.~95--108.

\bibitem{Lafont95}
\bysame, \emph{Equational reasoning for 2-dimensional diagrams}, Lecture Notes
  in Computer Science \textbf{909} (1995), 170--195.

\bibitem{Lafont03}
\bysame, \emph{Towards an algebraic theory of boolean circuits}, Journal of
  Pure and Applied Algebra \textbf{184} (2003), no.~2-3, 257--310.

\bibitem{Lafont06}
\bysame, \emph{Algebra and geometry of rewriting}, Preprint IML, 2006.

\bibitem{LafontMetayer06}
Yves Lafont and François Métayer, \emph{Polygraphic resolutions and homology of
  monoids}, Preprint IML, 2006.

\bibitem{Lankford79}
Dallas Lankford, \emph{On proving term rewriting systems are noetherian}, Tech.
  report, Louisiana Tech University, 1979.

\bibitem{Lawvere63}
Francis~William Lawvere, \emph{Functorial semantics of algebraic theories},
  Reprints in Theory and Applications of Categories \textbf{5} (2004), 1--121.

\bibitem{Metayer03}
François Métayer, \emph{Resolutions by polygraphs}, Theory and Applications of
  Categories \textbf{11} (2003), 148--184.

\bibitem{Plump99}
Detlef Plump, \emph{Term graph rewriting}, Handbook of Graph Grammars and
  Computing by Graph Transformation \textbf{2} (1999), 3--61.

\end{thebibliography}
\end{document}